# FauxCrypt - A Method of Text Obfuscation

Devlin M. Gualtieri
Consulting Scientist
Ledgewood, New Jersey
gualtieri@ieee.org

## Abstract

Warnings have been raised about the steady diminution of privacy. More and more personal information, such as that contained electronic mail, is moving to cloud computing servers where it might be machine-searched and indexed. FauxCrypt is an algorithm for modification of a plaintext document that leaves it generally readable by a person but not readily searched or indexed by machine. The algorithm employs a dictionary substitution of selected words, and an obfuscating transposition of letters in other words. The obfuscation is designed to leave the words understandable, although they are badly spelled. FauxCrypt is free, open source software, with source code available.

## Introduction

Cryptography is a process for converting plaintext to an unreadable cyphertext that can be recovered to plaintext through use of an alphanumeric key phrase and a cryptographic algorithm. The purpose of encryption is to make the plaintext unreadable by both people and machines in the absence of the key. Often, a writer's purpose is not to make the plaintext completely unreadable by people, but to modify it in a way to circumvent certain machine processes. This is not encryption, since a key is not used. It's obfuscation, since the plaintext can still be understood by a person, albeit with some difficulty. Examples of such obfuscation abound in the realm of email messaging, where authors of "spam" messages attempt to circumvent filters through creative misspelling. Their intent is to fool the machine but still get the message across to the person. Likewise, substitution of certain words for others (e.g., "pr0n") and slight modification of URLs ("hxxp" in place of "http") are used as a means to post messages that break local custom or rules on message boards.

Many people substitute phrases such as "*at*" in place of the "@" symbol in Internet postings to circumvent the robotic harvesting of email addresses. These modified email addresses are still decipherable by people without a special tool. The use of ROT13 encoding, a Caesar cypher in which letters of the 26 character English alphabet are substituted by letters 13 places distant (with modulus rotation to map "N" to "A", etc.), is often used to publish solutions to puzzles. A second application of ROT13 to a ROT13 encoded message recovers the plaintext without the need for a key. The intent of ROT13 is to make the plaintext easy to recover. Still, an operation is required for the recovery, akin to an encryption for which the key is commonly known, although a skilled person could train himself to read ROT13 encoded text on sight.

FauxCrypt has been developed as a way to modify plaintext in a way that makes it difficult for a machine to index, but simple for a person to read. FauxCrypt does not require a deciphering operation to make the text readable. The inspiration for this algorithm was an anonymous posting on the Internet in September 2003, purportedly from Cambridge University [1], that reads as follows:

"Aoccdrnig to a rscheearch at an Elingsh uinervtisy, it deosn't mttaer in waht oredr the ltteers in a wrod are, the olny iprmoetnt tihng is taht frist and lsat ltteer is at the rghit pclae. The rset can be a toatl mses and you can sitll raed it wouthit porbelm. Tihs is bcuseae we do not raed ervey lteter by it slef but the wrod as a wlohe."

This statement is still unattributed, and its origin is classified by some as an "urban legend" [2]. Whatever its source, it makes its point admirably in such a small number of words. FauxCrypt goes beyond the stated algorithm of just keeping the first and last letters of a word fixed. It uses certain processes to scramble the remainder of the letters in the word. The scrambling is done in a way that maintains reasonable readability.

Some previous programs obfuscate files in different ways. Balakrishnan, et al., have published a US patent application that modifies data files to eliminate private information by substitution [3]. This is quite unlike FauxCrypt, which does not remove content, but rather makes it more difficult, but not impossible, to interpret. The program, "strob," by Vedanta Barooah, converts text to a form not readily readable by machine through substitution of the HTML character codes for characters [4]. This makes text files considerably larger, but there is the advantage that the text is rendered as plaintext by web browsers.

## Algorithm

The FauxCrypt algorithm is summarized, as follows:

1) All characters are converted to lower case.
2) Words from a selected dictionary are replaced by equivalents, and these words are marked as not subsequently modifiable.
3) First and last letters of words are unchanged and these characters are marked as not subsequently modifiable.
4) Vowels in digraphs are swapped (oe -> eo) and these characters are marked as not subsequently modifiable.
5) The order of vowels and consonants in a word are kept the same, but some vowels are shifted forwards or backwards.
6) A single pair of consonants are swapped in words larger than N. Readability is presumed to be enhanced when a "riser" and a "dangler" are swapped; e.g., dg becomes gd).
7) For extreme obfuscation, a consonant or vowel can be moved quite a bit up or down.

Table I lists English digraphs and their frequency. Table II lists the "risers" and "danglers" of the English alphabet.

| Table I. Common English language digraphs and their frequency of occurrence [5]. (Source: http://www.math.cornell.edu/~mec/2003-2004/cryptography/subs/digraphs.html) | | | | | |
|---|---|---|---|---|---|
| Digraph | Frequency (%) | Digraph | Frequency (%) | Digraph | Frequency (%) |
| nt | 0.56 | hi | 0.46 | de | 0.09 |
| ha | 0.56 | is | 0.46 | se | 0.08 |
| es | 0.56 | or | 0.43 | le | 0.08 |
| st | 0.55 | ti | 0.34 | sa | 0.06 |
| en | 0.55 | as | 0.33 | si | 0.05 |
| ed | 0.53 | te | 0.27 | ar | 0.04 |
| to | 0.52 | et | 0.19 | ve | 0.04 |
| it | 0.5 | ng | 0.18 | ra | 0.04 |
| ou | 0.5 | of | 0.16 | ld | 0.02 |
| ea | 0.47 | al | 0.09 | ur | 0.02 |

| Table II. English language characters categorized as "risers" or "danglers." | |
|---|---|
| Risers | Danglers |
| b | g |
| d | j |
| f | p |
| h | q |
| k | y |
| l | |
| t | |

As an example of FauxCrypt operation, consider the first two paragraphs of Charles Dickens' "A Christmas Carol"[6], and their FauxCrypt output.

> "MARLEY was dead: to begin with. There is no doubt whatever about that. The register of his burial was signed by the clergyman, the clerk, the undertaker, and the chief mourner. Scrooge signed it: and Scrooge's name was good upon 'Change, for anything he chose to put his hand to. Old Marley was as dead as a door-nail.
>
> Mind! I don't mean to say that I know, of my own knowledge, what there is particularly dead about a door-nail. I might have been inclined, myself, to regard a coffin-nail as the deadest piece of ironmongery in the trade. But the wisdom of our ancestors is in the simile; and my unhallowed hands shall not disturb it, or the Country's done for. You will therefore permit me to repeat, emphatically, that Marley was as dead as a door-nail."

"marley was daed: to begin with. tehre is no duobt wahtever abuot taht. the regsiter of his burial was signed by the cyerglmna, the clerk, the udnertaker, and the cihef muorner. scrooge signed it: and scrooge's name was good upon 'cahgne, for anyhtnig he chose to put his hnad to. old marley was as daed as a doro-nail.

mnid! i don't maen to say taht i know, of my own knowgdele, waht tehre is paritcularly daed abuot a doro-nail. i mihgt have been inclnide, mfsely, to regard a cfofni-nail as the daedset piece of irnomnogery in the trade. but the wsidom of our ancseotrs is in the simile; and my unahllowed hnads sahll not dsiturb it, or the cuotnry's dnoe fro. you will teherfroe permit me to retaep, empahitcalyl, taht marley was as daed as a doro-nail."

**Analysis**

The Levenshtein distance [7] is a convenient measure of the difference between two strings. It's the minimum number of operations needed to transform one string into the other by deletion, insertion, or substitution of single characters. Since FauxCrypt does not insert characters into words, or delete characters from words, the Levenshtein distance will count substitutions, only. Another measure of string difference is a modified form of the Levenshtein distance developed by Damerau [8], known as the Damerau–Levenshtein distance. The Damerau–Levenshtein distance considers transposition of characters as a single operation, whereas such an operation would be doubly counted as substitutions by Levenshtein.

We analyzed the Levenshtein distance on a word-by-word basis between the entire Dickens' text of "A Christmas Carol" and its FauxCrypt translation. Letter case was ignored in this analysis. The 28,559 word texts had a total Levenshtein distance of 22,330, a Levenshtein distance of 0.782 per word, or roughly three-quarters of a character change per word. Because of the nature of the FauxCrypt algorithm, shorter words are generally unmodified, so the greatest contribution to the Levenshtein distance is born by longer words. Table III lists some of the longer Levenshtein distances between word pairs after FauxCrypt obfuscation. It can be seen that some words identified with a longer Levenshtein distance are less readable than others, but probably readable in their context.

| \_Table III\_. Some examples of FauxCrypt obfuscated words in Dickens' "A Christmas Carol," and their Levenshtein distance (LD) ||||||
|---|---|---|---|---|---|
| LD | Plaintext | FauxCrypted Text | LD | Plaintext | FauxCrypted Text |
| 5 | interesting | itnerseitng | 6 | miscellaneous | msicellnaeuos |
| 5 | confidential | cnofidneital | 6 | retirement | rteiermnet |
| 5 | alphabet | albahpte | 6 | preposterous | perpotseruos |
| 5 | undressing | udnerssnig | 7 | corporation | croprotaino |
| 5 | belonging | begnolnig | 7 | satisfactory | stasiyacotrf |
| 6 | conversations | cnoverstainos | 7 | endeavouring | ednaevuornig |

## Summary

FauxCrypt is a method for obfuscating text to make it difficult for machines to search and index, but still possible for people to read.

## Source Code Availability

Source code for FauxCrypt in the C programming language is listed in the appendix. FauxCrypt is free software, licensed under version 3 of the GNU General Public License, as published by the Free Software Foundation.

# Appendix – Source Code

```
/* -*- Mode: C; indent-tabs-mode: t; c-basic-offset: 4; tab-width: 4 -*- */

/***********************************************************************

FauxCrypt
Copyright (C) 2010 D. M. Gualtieri <gualtieri@ieee.org>
This version 23 April 2010

    This program is free software: you can redistribute it and/or modify
    it under the terms of the GNU General Public License as published by
    the Free Software Foundation, either version 3 of the License, or
    (at your option) any later version.

    This program is distributed in the hope that it will be useful,
    but WITHOUT ANY WARRANTY; without even the implied warranty of
    MERCHANTABILITY or FITNESS FOR A PARTICULAR PURPOSE.  See the
    GNU General Public License for more details.

    You should have received a copy of the GNU General Public License
    along with this program.  If not, see <http://www.gnu.org/licenses/>.

 ***********************************************************************/

/*
Algorithm for FauxCrypt:

1) Words from a selected dictionary are replaced by equivalents and marked
   as not subsequently modifiable. (e.g., the = teh; Not implemented in this
   version)
2) First and last letters of words are unchanged.
3) Vowels in digraphs are swapped (oe -> eo) and marked as not subsequently
   modifiable.
4) The order of vowels and consonants in a word are kept the same, but some
   vowels are shifted forwards or backwards.
5) A single pair of consonants are swapped in words larger than N.  This is
   best done when a "riser" and a "dangler" are shifted; e.g., dg becomes gd).
6) For extreme obfuscation, a consonant or vowel can be moved quite a bit
   up or down.  (Not implemented in this version)

Common english digraphs, implemented in this version: th he in er an re
nd at on nt ha es st en ed to it ou ea hi is or ti as te et ng of

Less common english digraphs, not implemented in this version:
al     de     se     le     sa     si     ar     ve     ra     ld     ur

Risers: b d f h k l t
Danglers: g j p q y

*/

#include <stdio.h>
#include <stdlib.h>

/* Prototypes */
long atol (const char *s);
char *strcpy (char *dest, const char *src);
void exit (int status);
void do_crypt (void);
void do_digraphs (void);
void do_danglers (void);
/* end of prototypes */
```

```c
long fctr;
int word_ctr;
int ch;
int temp_ch;
int this_word[32] = { 0x20 };
int fixed[32] = { 0 };

int i;
int position1;
int position2;
char fn1[64];
char fn2[64];
FILE *indata;
FILE *outdata;

void do_danglers (void)
{
//swap consonants
//Risers: b d f h k l t
//Danglers: g j p q y
//first check for presence

  position1 = 0;
  position2 = 0;

  for (i = 1; i < 30; i++)
//note loop limits - first and last characters are always unchanged
    {
      if (fixed[i] != 1)
        {
          if (this_word[i] == 'b')
            {
              position1 = i;
            }
          if (this_word[i] == 'd')
            {
              position1 = i;
            }
          if (this_word[i] == 'f')
            {
              position1 = i;
            }
          if (this_word[i] == 'h')
            {
              position1 = i;
            }
          if (this_word[i] == 'k')
            {
              position1 = i;
            }
          if (this_word[i] == 'l')
            {
              position1 = i;
            }
          if (this_word[i] == 't')
            {
              position1 = i;
            }
          if (this_word[i] == 'g')
            {
              position2 = i;
            }
          if (this_word[i] == 'j')
```

```
                 {
                    position2 = i;
                 }
                 if (this_word[i] == 'p')
                 {
                    position2 = i;
                 }
                 if (this_word[i] == 'q')
                 {
                    position2 = i;
                 }
                 if (this_word[i] == 'y')
                 {
                    position2 = i;
                 }

         }
      }
      if ((position1 != 0) && (position2 != 0))
      {
         temp_ch = this_word[position1];
         this_word[position1] = this_word[position2];
         this_word[position2] = temp_ch;
      }
}

void do_digraphs (void)
{
// Common english digraphs, implemented in this version: th he in er an re
// nd at on nt ha es st en ed to it ou ea hi is or ti as te et ng of

   for (i = 1; i < 30; i++)
//note loop limits - first and last characters are always unchanged
      {
         if ((fixed[i] != 1) && (fixed[i + 1] != 1))
         {
//th
            if ((this_word[i] == 't') && (this_word[i + 1] == 'h'))
            {
               (this_word[i] = 'h');
               (this_word[i + 1] = 't');
               fixed[i] = 1;
               fixed[i + 1] = 1;
            }
//he
            if ((this_word[i] == 'h') && (this_word[i + 1] == 'e'))
            {
               (this_word[i] = 'e');
               (this_word[i + 1] = 'h');
               fixed[i] = 1;
               fixed[i + 1] = 1;
            }
//in
            if ((this_word[i] == 'i') && (this_word[i + 1] == 'n'))
            {
               (this_word[i] = 'n');
               (this_word[i + 1] = 'i');
               fixed[i] = 1;
               fixed[i + 1] = 1;
            }
//er
            if ((this_word[i] == 'e') && (this_word[i + 1] == 'r'))
            {
```

```c
            (this_word[i] = 'r');
            (this_word[i + 1] = 'e');
            fixed[i] = 1;
            fixed[i + 1] = 1;
         }
//an
         if ((this_word[i] == 'a') && (this_word[i + 1] == 'n'))
         {
            (this_word[i] = 'n');
            (this_word[i + 1] = 'a');
            fixed[i] = 1;
            fixed[i + 1] = 1;
         }
//re
         if ((this_word[i] == 'r') && (this_word[i + 1] == 'e'))
         {
            (this_word[i] = 'e');
            (this_word[i + 1] = 'r');
            fixed[i] = 1;
            fixed[i + 1] = 1;
         }
//nd
         if ((this_word[i] == 'n') && (this_word[i + 1] == 'd'))
         {
            (this_word[i] = 'd');
            (this_word[i + 1] = 'n');
            fixed[i] = 1;
            fixed[i + 1] = 1;
         }
//at
         if ((this_word[i] == 'a') && (this_word[i + 1] == 't'))
         {
            (this_word[i] = 't');
            (this_word[i + 1] = 'a');
            fixed[i] = 1;
            fixed[i + 1] = 1;
         }
//on
         if ((this_word[i] == 'o') && (this_word[i + 1] == 'n'))
         {
            (this_word[i] = 'n');
            (this_word[i + 1] = 'o');
            fixed[i] = 1;
            fixed[i + 1] = 1;
         }
//nt
         if ((this_word[i] == 'n') && (this_word[i + 1] == 't'))
         {
            (this_word[i] = 't');
            (this_word[i + 1] = 'n');
            fixed[i] = 1;
            fixed[i + 1] = 1;
         }
//ha
         if ((this_word[i] == 'h') && (this_word[i + 1] == 'a'))
         {
            (this_word[i] = 'a');
            (this_word[i + 1] = 'h');
            fixed[i] = 1;
            fixed[i + 1] = 1;
         }
//es
         if ((this_word[i] == 'e') && (this_word[i + 1] == 's'))
         {
```

```
            (this_word[i] = 's');
            (this_word[i + 1] = 'e');
            fixed[i] = 1;
            fixed[i + 1] = 1;
          }
//st
        if ((this_word[i] == 's') && (this_word[i + 1] == 't'))
          {
            (this_word[i] = 't');
            (this_word[i + 1] = 's');
            fixed[i] = 1;
            fixed[i + 1] = 1;
          }
//en
        if ((this_word[i] == 'e') && (this_word[i + 1] == 'n'))
          {
            (this_word[i] = 'n');
            (this_word[i + 1] = 'e');
            fixed[i] = 1;
            fixed[i + 1] = 1;
          }
//ed
        if ((this_word[i] == 'e') && (this_word[i + 1] == 'd'))
          {
            (this_word[i] = 'd');
            (this_word[i + 1] = 'e');
            fixed[i] = 1;
            fixed[i + 1] = 1;
          }
//to
        if ((this_word[i] == 't') && (this_word[i + 1] == 'o'))
          {
            (this_word[i] = 'o');
            (this_word[i + 1] = 't');
            fixed[i] = 1;
            fixed[i + 1] = 1;
          }
//it
        if ((this_word[i] == 'i') && (this_word[i + 1] == 't'))
          {
            (this_word[i] = 't');
            (this_word[i + 1] = 'i');
            fixed[i] = 1;
            fixed[i + 1] = 1;
          }
//ou
        if ((this_word[i] == 'o') && (this_word[i + 1] == 'u'))
          {
            (this_word[i] = 'u');
            (this_word[i + 1] = 'o');
            fixed[i] = 1;
            fixed[i + 1] = 1;
          }
//ea
        if ((this_word[i] == 'e') && (this_word[i + 1] == 'a'))
          {
            (this_word[i] = 'a');
            (this_word[i + 1] = 'e');
            fixed[i] = 1;
            fixed[i + 1] = 1;
          }
//hi
        if ((this_word[i] == 'h') && (this_word[i + 1] == 'i'))
          {
```

```
              (this_word[i] = 'i');
              (this_word[i + 1] = 'h');
              fixed[i] = 1;
              fixed[i + 1] = 1;
           }
//is
        if ((this_word[i] == 'i') && (this_word[i + 1] == 's'))
           {
              (this_word[i] = 's');
              (this_word[i + 1] = 'i');
              fixed[i] = 1;
              fixed[i + 1] = 1;
           }
//or
        if ((this_word[i] == 'o') && (this_word[i + 1] == 'r'))
           {
              (this_word[i] = 'r');
              (this_word[i + 1] = 'o');
              fixed[i] = 1;
              fixed[i + 1] = 1;
           }
//ti
        if ((this_word[i] == 't') && (this_word[i + 1] == 'i'))
           {
              (this_word[i] = 'i');
              (this_word[i + 1] = 't');
              fixed[i] = 1;
              fixed[i + 1] = 1;
           }
//as
        if ((this_word[i] == 'a') && (this_word[i + 1] == 's'))
           {
              (this_word[i] = 's');
              (this_word[i + 1] = 'a');
              fixed[i] = 1;
              fixed[i + 1] = 1;
           }
//te
        if ((this_word[i] == 't') && (this_word[i + 1] == 'e'))
           {
              (this_word[i] = 'e');
              (this_word[i + 1] = 't');
              fixed[i] = 1;
              fixed[i + 1] = 1;
           }
//et
        if ((this_word[i] == 'e') && (this_word[i + 1] == 't'))
           {
              (this_word[i] = 't');
              (this_word[i + 1] = 'e');
              fixed[i] = 1;
              fixed[i + 1] = 1;
           }
//ng
        if ((this_word[i] == 'n') && (this_word[i + 1] == 'g'))
           {
              (this_word[i] = 'g');
              (this_word[i + 1] = 'n');
              fixed[i] = 1;
              fixed[i + 1] = 1;
           }
//of
        if ((this_word[i] == 'o') && (this_word[i + 1] == 'f'))
           {
```

```c
              (this_word[i] = 'f');
              (this_word[i + 1] = 'o');
              fixed[i] = 1;
              fixed[i + 1] = 1;
            }

        }

    }

}

void do_crypt (void)
{
  if (word_ctr == 0)
    {
//nothing to print
      return;
    }

  if ((word_ctr == 1) || (word_ctr == 2))
    {
//no changes for single or double character words
      for (i = 0; i < (word_ctr); i++)
        {
          fprintf (outdata, "%c", this_word[i]);
        }
      return;
    }

  if (word_ctr > 2)
    {
//do some changes before printing
      do_digraphs ();
      do_danglers ();
      for (i = 0; i < (word_ctr); i++)
        {
          fprintf (outdata, "%c", this_word[i]);
        }
      return;
    }

}
int main (int argc, char *argv[])
{
  if (argc < 2)
    {
      printf ("Usage: fauxcrypt input.txt output.txt\n");
      exit (1);
    }

  strcpy (fn1, argv[1]);
  printf ("\nInput file selected = %s\n", fn1);

  if ((indata = fopen (fn1, "rb")) == NULL)
    {
      printf ("\nInput file cannot be opened.\n");
      exit (1);
```

```c
      }
    strcpy (fn2, argv[2]);
    printf ("\nOutput file selected = %s\n", fn2);

    if ((outdata = fopen (fn2, "w")) == NULL)
      {
        printf ("\nOutput file cannot be opened.\n");
        exit (1);
      }

/* ASCII Codes
Numbers 0-9 are 0x30 through 0x39
Upper case A-Z are 0x41 through 0x5A
Lower case a-z are 0x61 through 0x7A
*/

    fctr = 0;
    word_ctr = 0;
    ch = 0;
//first position is always fixed
    fixed[0] = 1;

    while (ch != EOF)
      {
        ch = fgetc (indata);
        fctr++;

//convert to lower case
        if ((ch > 0x40) && (ch < 0x5B))
          {
            ch = ch + 0x20;
          }

//parse and store word
        if (((ch > 0x60) && (ch < 0x7B)) || ((ch > 0x2A) && (ch < 0x3A)))
          {
            this_word[word_ctr] = ch;
            word_ctr++;
          }
        else
          {
//last character always fixed
            fixed[word_ctr - 1] = 1;
            do_crypt ();
//print punctuation
            fprintf (outdata, "%c", ch);

            word_ctr = 0;
            for (i = 0; i < 32; i++)
              {
//set all character positions as non-fixed initially
                fixed[i] = 0;
//set word as all blanks initially
                this_word[i] = 0x20;
              }
//first position is always fixed
            fixed[0] = 1;
          }

        if (fctr % 1000 == 0)
          printf ("*");

      }
```

```
    fclose (indata);
    fclose (outdata);
    printf ("\nDone.\n");

    return 0;

}
```